\documentclass[journal,comsoc]{IEEEtran}
\usepackage{color}
\usepackage{lettrine}
\usepackage{xcolor}
\usepackage{makecell}
\usepackage{graphicx}
\usepackage{float}
\usepackage{booktabs}
\usepackage{cite}

\begin{document}
\title{Extreme-MIMO Field Trials in 7~GHz Band: \\Unlocking the Potential of New Spectrum for 6G}

\author{Seunghyun Lee, Jungmin Yoon, Sangwon Jung, Young-Han Nam,~\textit{Member, IEEE}, Gary Xu,~\textit{Member, IEEE},\\ Chan-Byoung Chae,~\textit{Fellow, IEEE}, Juho Lee,~\textit{Fellow, IEEE}, and Jianzhong (Charlie) Zhang,~\textit{Fellow, IEEE}
\thanks{S. Lee, J. Yoon, S. Jung, and J. Lee are with Samsung Research, Samsung Electronics, Seoul 06765, Korea (e-mails: {sh0108.lee, jm.yoon, sangwon.jung, juho95.lee}@samsung.com). 
}
\thanks{Y.-H. Nam, G. Xu, and J. Zhang are with Samsung Research America, Plano, TX 75024, USA (e-mails: {younghan.n, gary.xu,  jianzhong.z}@samsung.com)}
\thanks{C.-B. Chae is with the School of Integrated Technology, Yonsei University, Seoul 03722, Korea (e-mail: cbchae@yonsei.ac.kr).}
}

\maketitle

\begin{abstract}
The frequency range around 7 GHz has emerged as a promising upper mid-band spectrum for 6th generation (6G), offering a practical balance between coverage and capacity. To fully exploit this band, however, future systems require substantially stronger beamforming and spatial multiplexing capability than today's 5G 64-port commercial deployments. This article investigates extreme multiple-input multiple-output (X-MIMO) with 256 digital ports as a practical 6G architecture for 7 GHz operation. First, through system-level simulations, we examine the throughput benefits and design trade-offs of increasing the number of base station (BS) and user equipment (UE) digital antenna ports, including comparisons between 128-port and 256-port configurations. We then present a 256-port 7 GHz BS and UE prototype and report field-trial results obtained in urban outdoor environments. The measurements demonstrate the feasibility of 8-layer downlink single-user MIMO over a 100 MHz bandwidth, achieving more than 3 Gbps for a single user under urban outdoor propagation conditions. Channel analysis based on measured data further suggests how the large digital aperture of X-MIMO supports high-order spatial multiplexing even with limited dominant angular clusters. Finally, we identify key challenges and outline research directions toward practical deployment of 7 GHz X-MIMO systems for 6G.

\end{abstract}

\section{Introduction}
\label{sec:intro}

\IEEEPARstart{T}{he} commercial success of 5th generation (5G) has been largely driven by the utilization of the C-band (3.5-4.8 GHz), which offered a significant leap in capacity over legacy long-term evolution (LTE) bands. However, as we look toward the 2030s and the era of 6G, the industry faces two major challenges: (1) the exhaustion of available C-band spectrum and (2) the emergence of hyper-intensive applications such as artificial intelligence (AI)-based services, digital twins, and immersive extended reality (XR)~\cite{c1}.

The frequency range around 7~GHz is increasingly regarded as a ``golden band" for 6G, offering a strategic balance between the wide-area coverage of lower frequencies and the massive bandwidths associated with millimeter-wave (mmWave) bands \cite{c2}-\cite{c6}. The global momentum toward securing this band for 6G is accelerating, driven by regulatory bodies and governments worldwide. During the World Radiocommunication Conference 2023 (WRC-23), portions of the 6.425--7.125 GHz band were identified for international mobile telecommunications (IMT) in several regions \cite{c2}, and looking ahead, the 7.125--8.4 GHz range has been proposed as a key agenda item for WRC-27. Moreover, national governments are taking proactive steps. For example, the U.S. government recently issued a memorandum aimed at repurposing the 7.125 GHz band for mobile use \cite{c3}. Similarly, initiatives in Europe (targeting the upper part of the 6 GHz range \cite{c4}) and South Korea (targeting 7.125--8.4 GHz \cite{c5}) are actively exploring the upper mid-band as a primary candidate for 6G services. These global trends suggest that a significant amount of contiguous bandwidth--ranging from 200 MHz up to 400 MHz--could eventually be allocated to individual mobile network operators. Reflecting this momentum, 3rd Generation Partnership Project (3GPP) has already initiated discussions on 6G specifications, with a focus on supporting channel bandwidths of up to 400 MHz \cite{c6}.

To fully exploit the potential of the 7 GHz band, a fundamental advancement in multiple-input multiple-output (MIMO) technology is required. While 5G revolutionized wide-area capacity with 64T64R (64 transmit and receive antenna ports, i.e., 64-port) massive MIMO radios \cite{c7}, 6G demands more aggressive spatial multiplexing and beamforming strategies to overcome higher path loss and maximize spectral efficiency.  In this article, we propose and validate extreme-MIMO (X-MIMO) technology. Specifically, by leveraging the shorter wavelength at 7 GHz, we demonstrate how 256T256R (i.e., 256-port) digital architectures with a significantly larger number of antenna elements can be integrated into existing base station (BS) form factors. This architecture enables a substantial improvement in spectral efficiency and supports higher-order spatial multiplexing for next-generation devices.


 \begin{figure*}[t]
\centering
\includegraphics[width=\textwidth]{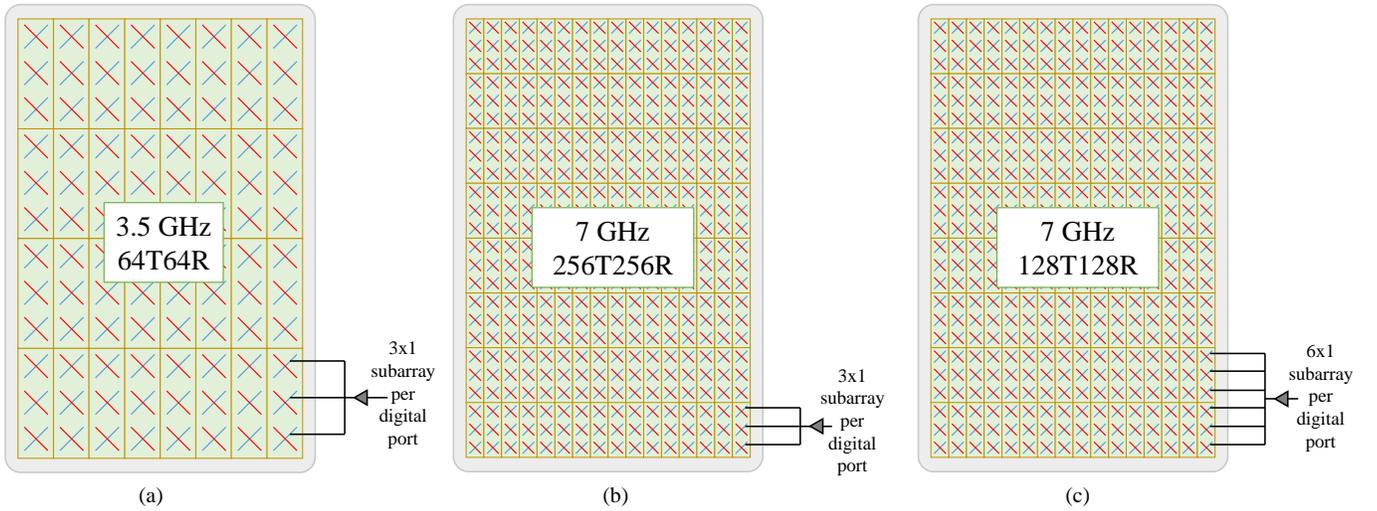}
\caption{Comparison of RU antenna architectures for 5G C-band (e.g., 3.5 GHz) and 6G at 7 GHz. For 6G at 7 GHz, two antenna architecture options are also presented: 256T256R and 128T128R.}
\label{fig:Fig1}
\end{figure*}


In this article, we connect the architectural potential of 7~GHz X-MIMO to practical system evidence. We first provide comprehensive analysis using extensive system-level simulation (SLS) results to quantify the capacity gains of 256-port X-MIMO in both single-user (SU) and multi-user (MU) MIMO scenarios. We then examine architectural trade-offs in BS antenna frontend design options for 7 GHz X-MIMO systems. Next, we report field validation results using a custom-built 256-port 7 GHz X-MIMO testbed platform, demonstrating that high-order spatial multiplexing with eight data streams is achievable in practical urban cellular environments. The field measurement results are further supported by in-depth channel analysis, revealing the potential of X-MIMO with significantly enhanced spatial resolution and digital beamforming capabilities. Finally, we outline promising future research directions that provide useful insights into the standardization and commercialization of extreme-scale MIMO systems for 6G.


The importance of this architectural scaling is not limited to recovering the additional path loss at 7 GHz. More fundamentally, the 7 GHz band creates a unique operating point at which dense antenna integration, relatively wide-area coverage, and meaningful multi-layer MIMO transmission can coexist within a practical cellular form factor. In lower bands, the physical array size becomes a limiting factor, whereas in much higher bands coverage and blockage become dominant concerns. The upper mid-band around 7 GHz therefore represents a particularly attractive regime in which a substantially larger digital array can be realized while still preserving deployment characteristics that resemble today's macro and micro cellular networks. This is precisely why X-MIMO should be viewed not as an incremental extension of 5G massive MIMO, but as a new architectural opportunity enabled by the 7 GHz spectrum itself.

\section{7 GHz X-MIMO: Opportunities for Extreme Capacity}
\label{sec:explain}
The transition toward 6G requires new spectrum that can simultaneously support wide-area coverage and extreme capacity. In today's 5G commercial landscape, 64T64R massive MIMO radio units (RUs), i.e., those utilizing 64 digital ports, serve as the workhorse of the mid-band market, particularly in the C-band around 3.5 GHz \cite{c7}. The shift to the 7 GHz band presents a transformative opportunity to further enhance spatial multiplexing, which lies at the core of the X-MIMO concept.

As illustrated in Fig. 1(a), a typical 5G 3.5 GHz 64T64R RU connects each of its 64 digital ports to a 3~$\times$~1 subarray of cross-polarized antenna elements, resulting in a total of 192 antenna elements. Since the wavelength at 7 GHz is approximately half that at 3.5 GHz, antenna elements can be packed four times more densely within the same RU footprint. This enables a 256T256R X-MIMO architecture (see Fig. 1(b)) with 768 antenna elements. This configuration provides significantly higher equivalent isotropic radiated power (EIRP) and finer spatial resolution, enabling narrow and precise beamforming suitable for operation at 7~GHz.


Alternatively, a 128T128R design (see Fig. 1(c)) can be implemented within the same physical dimensions by using 6~$\times$~1 subarrays. Although this 128-port solution offers a more cost-effective profile with reduced power consumption and baseband complexity while maintaining the high EIRP, it involves a trade-off in digital beamforming granularity and spatial multiplexing performance.

In the following two subsections, we explore the capacity potential of this new spectrum band and provide a trade-off analysis comparing the 128T128R and 256T256R solutions in both SU-MIMO and MU-MIMO scenarios.

\subsection{Opportunities for SU-MIMO Capacity Expansion in 7 GHz}

The hardware integration advantages of 7 GHz, due to its shorter wavelength compared to 3.5 GHz, also extend to the user equipment (UE). In 5G C-band implementations, smartphone antenna configurations are typically limited to a maximum of four receive (Rx) antennas, primarily due to physical form factor constraints \cite{c8}, effectively capping downlink spatial multiplexing at four data streams (i.e., four MIMO layers). At 7 GHz, the shorter wavelength allows mobile devices to accommodate a larger number of antennas, such as eight Rx ports (i.e., 8-Rx), without increasing the overall device size.


It is also worth noting that initial deployments for more than 4-Rx are likely to find a natural home in fixed wireless access (FWA) applications. Since customer premises equipment (CPE) typically has more flexible size constraints than mobile handsets \cite{c8}, it serves as an ideal platform to exploit high-order MIMO data streams and demonstrate the massive throughput gains possible in this new spectrum. Together with the smartphone applications, the opportunities with having more than four receive antennas can be explored, thereby significantly enhancing both capacity and coverage performance with X-MIMO.


\begin{table}[t]
\centering
\caption{SU-MIMO Simulation Results}
\begin{tabular}{cccc}
\toprule
\makecell{UE Rx\\ports} &
\makecell{Avg. of Relative \\  UE Throughput} &
\makecell{5\%-tile of Relative \\ UE Throughput} &
\makecell{Avg. \# of \\ Scheduled Layers} \\
\midrule
4 & 100\% & 100\% & 3.44 \\
8 & 135\% & 154\% & 4.46 \\
\bottomrule
\end{tabular}
\end{table}

To verify these benefits, we conducted SLS based on typical 3GPP evaluation methodologies \cite{c9}. Assuming a 256-port BS in Fig. 1(b), we compared UEs equipped with 4-Rx and 8-Rx in a scenario with 10 UEs per cell, 500-meter inter-site distance (ISD), full-buffer downlink traffic, and the urban macro (UMa) channel model environment. The results, summarized in Table 1, show that adopting 8-Rx at UEs leads to a 35\% increase in average UE throughput and a 54\% gain in 5th-percentile UE (i.e., cell-edge UE) throughput in the downlink, compared to the 4-Rx case. Furthermore, based on the typical link adaptation and scheduling mechanisms, the 256-port X-MIMO system with 8-Rx UEs successfully enables downlink transmissions with more than four layers in over 30\% of total downlink transmission occasions, resulting in an average of 4.46 scheduling layers in the downlink. These results indicate the strong capacity potential of the 7 GHz band for extremely high data-rate services when larger BS and UE antenna configurations can be supported.

An important observation from Table I is that the gain from 8-Rx is even larger at the 5th-percentile than at the average-user level. This behavior is meaningful from a deployment perspective because it indicates that the additional UE-side receive dimensions are useful not only for peak-rate enhancement, but also for improving robustness in weaker channel conditions. Intuitively, when the UE can resolve and combine more spatial components, the BS has greater flexibility in selecting transmission layers and precoding directions that remain effective even near the cell edge. As a result, the benefit of increasing UE receive capability extends beyond simple peak multiplexing gain and also contributes to the improvement in cell coverage.

In Section~\MakeUppercase{\romannumeral 4}, we present field-trial results obtained with our 256-port X-MIMO prototype. The measurements demonstrate the feasibility of over-the-air 8-layer downlink transmission at 7 GHz in practical urban outdoor environments, enabled by the enhanced spatial multiplexing and digital beamforming capability of the large-port architecture.

\begin{figure}[t]
\centering
\includegraphics[width=0.5\textwidth]{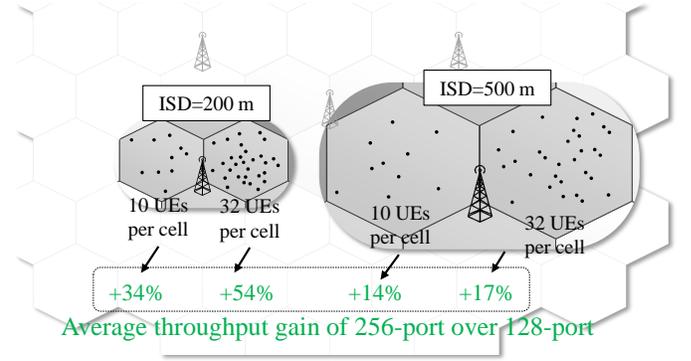}
\caption{X-MIMO system-level simulation results for MU-MIMO under two ISD scenarios (200 m and 500 m), with two user densities of 10 and 32 UEs per cell.}
\label{fig:Fig2}
\end{figure}

\subsection{Enhanced MU-MIMO Capability with X-MIMO}

Beyond SU-MIMO gains, increasing the number of digital ports at the BS drastically enhances MU-MIMO capabilities through superior spatial resolution. To verify this and obtain technical insights, in this subsection we compare 128-port and 256-port systems in Figs. 1(c) and 1(b), respectively, based on conventional 3GPP SLS methodology for MU-MIMO \cite{c9} across two distinct ISD scenarios (i.e., 200 meters and 500 meters) and two user densities (i.e., 10 and 32 UEs per cell). Similar to the SLS in Section \MakeUppercase{\romannumeral 2}-A, full-buffer downlink traffic and UMa channel model are assumed.

As shown in Fig. 2, the 256-port system significantly outperforms the 128-port system in the 200 meters ISD scenario. This smaller ISD scenario results in the higher SNR environment, which increases the probability of pairing more UEs for MU-MIMO in typical greedy-based MU pairing algorithms, where the 256-port BS can support more simultaneous users with better spatial resolution and interference suppression capabilities over the 128-port BS. Interestingly, the performance gain of the 256-port system becomes even more pronounced as the cell becomes ``crowded" (32 UEs per cell), as the BS more actively utilizes its spatial degrees of freedom. This implies that for dense-urban deployments in megacities like Seoul, the 256-port system offers a clear advantage. Conversely, the 128-port system remains a viable, cost-effective solution for larger ISD environments (e.g., urban or rural macro cells), where the throughput performance gap compared to the 256-port system is smaller, as can be observed from the result with ISD = 500 m in Fig. 2.

From a network-design viewpoint, these results suggest that 128-port and 256-port architectures should not be viewed as competing in a strictly binary sense, but rather as addressing different operating points. A 256-port design is especially attractive in dense urban deployments, hotspot layers, and traffic-heavy upper mid-band cells where the scheduler can frequently exploit high SNR and multi-user separation opportunities. By contrast, a 128-port configuration may remain a practical choice for less dense deployments or coverage-oriented scenarios where the incremental MU-MIMO gain is smaller and implementation cost becomes a stronger consideration. This architectural perspective is important for early 6G rollout strategies, where operators may adopt different X-MIMO scales depending on site density, traffic demand, and fronthaul/baseband constraints.

\section{7 GHz X-MIMO Testbed Platforms}
\label{sec:test}

To validate the theoretical advantages of 7 GHz X-MIMO presented in Section \MakeUppercase{\romannumeral 2}, we developed an end-to-end 7 GHz X-MIMO testbed including both the BS and UE (see Fig. 3). The testbed was designed to evaluate real-world performance in representative 6G deployment scenarios through field trials.


\begin{figure}[t]
\centering
\includegraphics[width=0.5\textwidth]{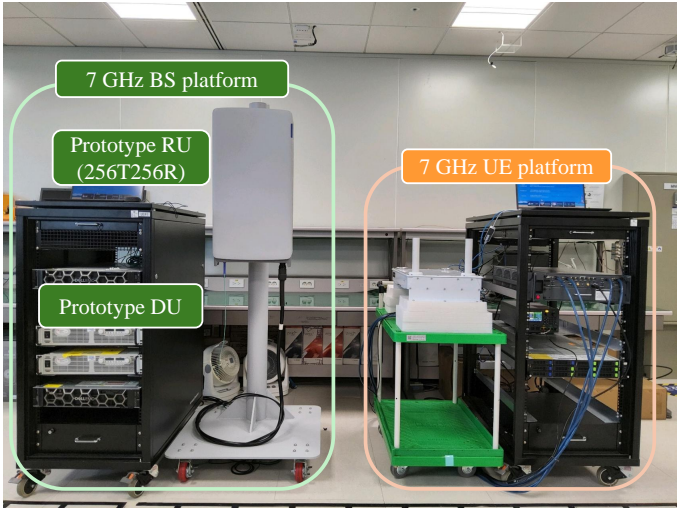}
\caption{X-MIMO testbed showing the BS and UE platforms for 7 GHz field tests.}
\label{fig:Fig3}
\end{figure}
\subsection{7 GHz BS Testbed Platform} 
The 7 GHz BS prototype operates in the 7.125 GHz band with a 100 MHz channel bandwidth. The BS platform consists of a prototype RU and digital unit (DU). The RU is equipped with 256 digital ports, enabling high-order MIMO transmission and reception. The platform is developed using a customized 6G pre-standard air-interface specifications, supporting basic legacy 5G functionalities as well as advanced features such as 256-port channel state information reference signal (CSI-RS) transmission, 8-layer downlink SU-MIMO, and the real-time baseband processing required for X-MIMO signal processing. From an implementation perspective, the prototype requires careful integration of large-scale radio frequency (RF) chains, synchronization across many digital ports, high-capacity fronthaul interface, and a practical real-time processing pipeline. These challenges become significantly more demanding as the port count scales from today's commercial 64-port systems to 256 ports.


\subsection{7 GHz UE Prototype}
\begin{figure*}[t]
\centering
\includegraphics[width=\textwidth]{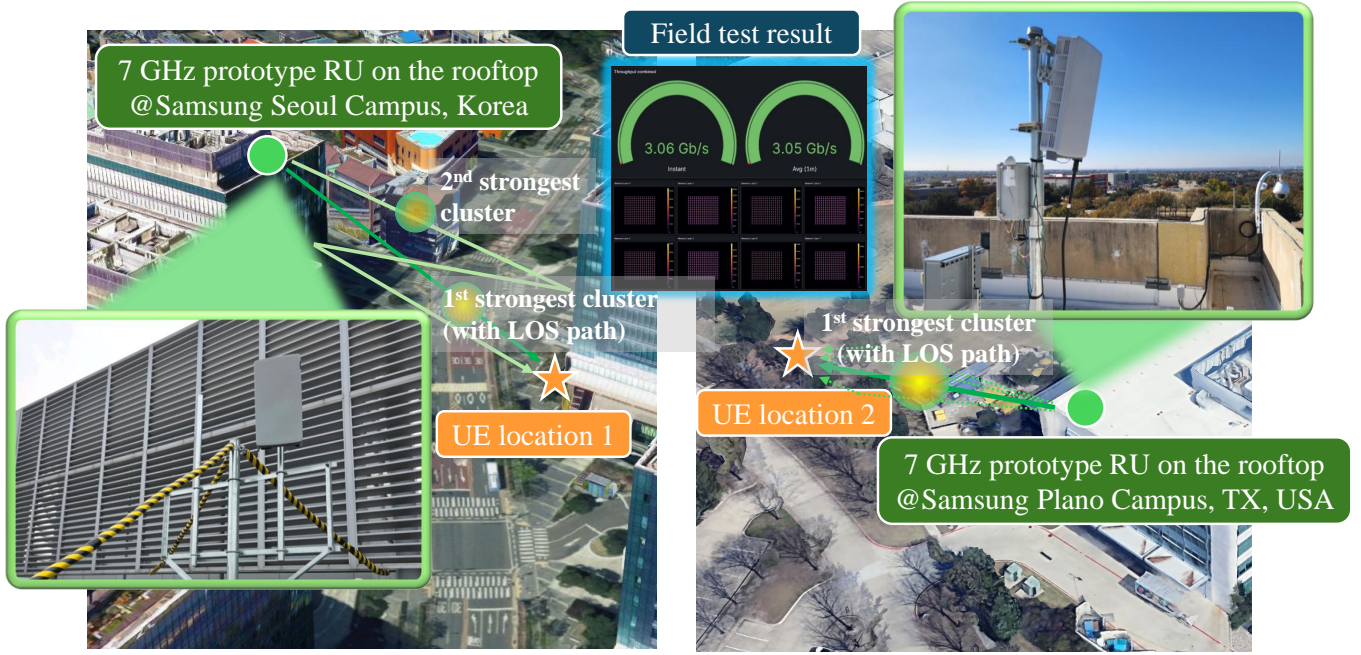}
\caption{Field test setup, where the 7 GHz prototype BS is mounted on rooftops at the Samsung Seoul Campus, South Korea (left) and Samsung Plano Campus, TX, USA (right). Two UE locations are also indicated, where 8-layer peak data rates were achieved, along with the corresponding dominant signal-path clusters. The measured field-test result of 3.05 Gbps (averaged over one minute) and the corresponding constellation maps for 8-layer with 256QAM are presented, captured at the UE.}
\label{fig:Fig4}
\end{figure*}

The UE prototype is developed in collaboration with Keysight Technologies.
To overcome the conventional four-layer limit, the prototype is equipped with an integrated array of 8 Rx ports. This hardware enhancement enables full 8-layer downlink MIMO reception, serving as an important proof-of-feasibility platform for future 6G device evolution. The platform also supports customized 6G pre-standards air interface specifications, enabling seamless integration with the BS testbed introduced in Section~III-A. The prototype should therefore be viewed as a research platform for validating high-order MIMO feasibility at 7 GHz, rather than as a commercial UE form factor. Nevertheless, it provides important insights into how larger UE-side antenna configurations can be exploited in future 6G devices, particularly in early FWA and other size-tolerant use cases.


\subsection{End-to-End System Integration and Real-Time Validation}
The integrated BS and UE platforms establish an end-to-end validation environment for 7 GHz X-MIMO beyond a simple laboratory setup. They enable system-level evaluation and field data collection of key operations, including CSI acquisition, digital precoding, synchronization, layer adaptation, and real-time processing, under realistic propagation conditions. This capability is particularly critical for 7 GHz X-MIMO, where performance gains depend not only on array size but also on whether the associated control and baseband procedures can be realized in real time at scale in this new frequency band.


\section{Field Test Results and Analysis}
\label{sec:result}
To validate the real-world viability of the system, we conducted an extensive field-testing campaign focusing on capacity gains from high-order spatial multiplexing. In this section, we present the field test setup and results for downlink capacity, along with a technical analysis of the measured MIMO channels based on the collected dataset.


\subsection{Field Test Setup}

As shown in Fig. 4, the experimental campaign was conducted at two sites: Samsung Seoul R\&D Campus in Seoul, Korea, and Samsung Plano Campus in Texas, USA. The 7 GHz prototype BS RU was installed on rooftops at both sites to emulate typical urban macro- and micro-cell deployments in commercial cellular networks. The downlink tests were carried out using measurements collected with the 8-Rx UE prototype at locations up to 300 meters from the BS, representing urban outdoor-to-outdoor scenarios.


The field campaign was designed with two complementary objectives. First, it aimed to verify whether the new frequency of 7 GHz and the 256-port architecture could support stable high-order downlink transmission in real outdoor propagation environments, rather than only in simulations or lab emulations. Second, it aimed to collect measured channel data to help explain under what conditions high-order multiplexing becomes feasible, and why. Accordingly, while the main throughput results reported in this article correspond to selected representative UE locations, the broader measurement campaign also served to characterize channel behavior of the 7 GHz bands over a wider outdoor range of up to approximately 300 m.


\subsection{Capacity Test Results and Field Channel Analysis}
The field trials demonstrate that high-order spatial multiplexing is practically achievable at 7 GHz using the proposed X-MIMO architecture. Using the 8-Rx UE prototype, we achieved stable 8-layer downlink SU-MIMO transmission at two representative UE locations, both approximately 100 m from the BS, as shown in Fig. 4. Under these favorable outdoor conditions, where a clear line-of-sight (LOS) path was maintained between the BS and UE, the measured downlink throughput exceeded 3 Gbps for a single user over a 100~MHz channel with 256QAM. These results should therefore be interpreted as a proof of feasibility in selected urban outdoor scenarios, rather than as a universal performance guarantee across all deployment conditions.

Nevertheless, the measured throughput level is notable in its own right. Achieving more than 3 Gbps for a single user within only 100 MHz of bandwidth implies a very high realized spectral-efficiency regime, enabled jointly by 8-layer spatial multiplexing, high-order modulation, and precise large-array beamforming. In other words, the result is not merely a consequence of wide bandwidth, but a direct demonstration that the spatial degrees of freedom offered by the 256-port architecture can be translated into substantial user-plane capacity under real propagation conditions in the 7 GHz band. This observation is especially relevant for 6G because it suggests that upper mid-band systems may deliver fiber-like user throughput without relying exclusively on mmWave-scale bandwidth.



To better understand how the measured channels supported this result, we further analyze the field-measured CSI-RS data collected by the UE prototype. This analysis provides a physical interpretation of the observed multiplexing behavior and helps explain how the large digital aperture of the 256-port BS can exploit the spatial structure of the propagation channel. These data are obtained from the received signals at the UE prototype, specifically using pre-standard 256-port CSI-RS customized for our X-MIMO testbed. As shown in Fig. 5, we transform the estimated wireless channel derived from the measured CSI-RS into power-angular profile heatmaps for each of the two UE locations presented in Fig. 4. These heatmaps represent the relative strength of signal paths across the azimuth angle of departure (AoD) and zenith angle of departure (ZoD). In each heatmap in Fig. 5, we also highlight eight orthogonal discrete Fourier transform (DFT) directions obtained from the digital precoding matrix at the BS, each corresponding to one of the eight transmission layers. The key observations for each UE location are summarized as follows:

1)	UE location 1 (Fig. 5(a)): The measured angular profile suggests that the propagation environment at UE location 1 contains at least two dominant clusters. By comparing the physical direction of the actual LOS path and the AoD/ZoD of the strongest cluster (obtained from the power-angular profile in Fig. 5(a)), we conjecture that this strongest cluster contains the LOS path and other nearby paths owing to the reflectors around the UE. Similarly, by analyzing the AoD/ZoD (obtained from the power-angular profile in Fig. 5(a)) as well as the propagation delay (obtained from the actual power-delay profile) of the paths in the second strongest cluster, we interpret that the second strongest cluster comprises the paths experiencing double reflections from buildings (see Fig. 4). It is observed that five data streams (i.e., Layers 1, 2, 5, 6, and 8) are multiplexed in the spatial domain over the strongest cluster, while the other three data streams (i.e., Layers 3, 4, and 7) are multiplexed over the second-strongest cluster. 

2)	UE location 2 (Fig. 5(b)): The measured angular profile is dominated by a single strongest cluster. As in UE location~1, this dominant cluster is consistent with the LOS direction. Interestingly, it is observed that only a single cluster (i.e., the strongest cluster) is utilized for multiplexing all the eight data streams. It should be noted that the propagation environment at UE location 2 was characterized by the presence of nearby trees. Thus, it is inferred that the scattering effects caused by foliage in fact influenced the channel characteristics, contributing additional intra-cluster multipath richness and spatial diversity within the strongest cluster.

It is worth noting that multiple data streams can be orthogonally multiplexed over only a small number of physical channel clusters, or even a single cluster, which is an interesting observation from the field measurements. This is enabled by X-MIMO equipped with a large number of digital ports (i.e., 256 ports), providing extremely fine spatial resolution and enabling the formation of narrow, highly directional beams for each digital precoding vector. This highlights a key advantage of our 7 GHz X-MIMO system: its enhanced digital precoding and beamforming capabilities enable high-order spatial multiplexing even in environments with limited angular spread or sparse scattering.

While the measurement results in this section focus on SU-MIMO, mainly due to the practical difficulty of preparing multiple synchronized UE prototypes for field testing, the implications extend beyond the single-user case. The same large-port architecture that enables high-order SU-MIMO also provides finer spatial discrimination across users, which is the key ingredient for MU-MIMO scheduling and inter-user interference suppression. Therefore, the field results reported here should be interpreted as direct evidence that the array has sufficient spatial resolution to support very high-dimensional transmission in practice. Combined with the system-level results in Section II-B, this strengthens the case that 7 GHz X-MIMO can also provide substantial MU-MIMO gains in dense deployment scenarios, although full multi-user field validation remains an important next step.

\begin{figure}[t]
\centering
\includegraphics[width=0.5\textwidth]{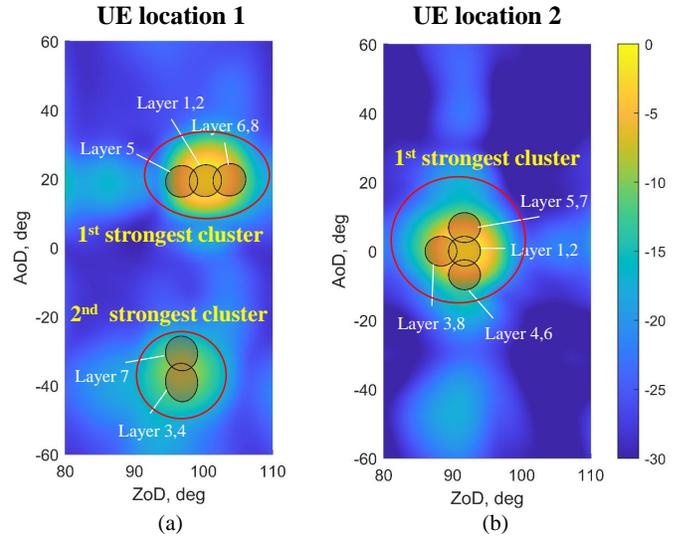}
\caption{Power-angular profile heatmaps and DFT beams used for 8-layer multiplexing at two UE locations across two measurement sites: (a) Seoul and (b) Plano, where a peak downlink data rate of 3 Gbps is achieved with 8-layer SU-MIMO transmission.}
\label{fig:Fig5}
\end{figure}
\section{Future Research Directions}
\label{sec:future}

The results presented in this article establish the feasibility of 7 GHz X-MIMO in selected outdoor field conditions, but several critical challenges remain before such systems can be deployed at scale. In particular, broader validation is still needed for coverage-limited scenarios, uplink implementation, and operational complexity in large-port systems. In this section, we highlight research directions that arise directly from the lessons learned through our simulations, prototype development, and field trials.


\subsection{Coverage Enhancement for X-MIMO}

Maintaining a robust link budget remains a primary challenge due to the inherent path loss of the 7 GHz band. This is particularly critical in outdoor-to-indoor scenarios, where signals face additional attenuation from modern building materials. Future research directions for narrowing the coverage gap relative to lower 5G bands include high-resolution beamforming and cell-specific digital beamforming techniques leveraging AI and integrated sensing and communications (ISAC)~\cite{c10}. Furthermore, new advanced waveform design for lower peak-to-average power ratio (PAPR) characteristics with e.g., frequency-domain spectrum-shaping (FDSS) \cite{c11} is also a promising area of research and field validation, to enhance the cell coverage owing to maximizing the power amplifier (PA) efficiency at both BS and UE.

In particular, several coverage-related questions remain open from a field-validation standpoint. These include outdoor-to-indoor penetration loss, mobility-induced beam tracking stability, and performance at larger cell radii where both path loss and angular resolution become more challenging. Another important issue is how much of the X-MIMO gain observed in favorable outdoor locations can be preserved under partial blockage or richer but weaker scattering environments. Addressing these questions will require not only link-level innovation but also broader and more systematic field campaigns across diverse deployment conditions.

\subsection{Uplink Enhancement for X-MIMO}
The 256-port architecture in X-MIMO offers significant benefits for uplink reception. The Open Radio Access Network (O-RAN)-based 7-2x split, widely adopted in commercial 5G massive MIMO BSs, provides a favorable trade-off between communication performance and fronthaul overhead. Recent standard updates on uplink demodulation reference signal (DMRS)-based beamforming further improve uplink capacity under highly interference-limited conditions. For X-MIMO, however, 128- or 256-port digital chains introduce additional challenges in achieving an optimal trade-off among performance, complexity, and fronthaul overhead in 3GPP and O-RAN standards, as well as in DU and RU implementations. Promising solutions include digital port compression (also referred to as port reduction) leveraging advanced receive digital beamforming based on UE CSI feedback or uplink DMRS, AI-driven scalable channel estimation, and advanced MIMO receiver designs. Moreover, it is also worth investigating the potential performance gains of uplink MIMO with coherent digital beamforming and spatial multiplexing at the UE side. By exploiting the opportunity to deploy more antenna ports at the UE, enabled by the shorter wavelength at 7 GHz, both theoretical advancements and corresponding field trials of advanced uplink MIMO constitute a promising research direction for future 6G commercialization.

Uplink design deserves separate attention because its bottlenecks are not identical to those of downlink transmission. In downlink, the main question is whether the BS can exploit a large aperture to form sharp beams and multiplex many layers. In uplink, however, the receiver must additionally cope with scalable channel estimation, real-time combining across a very large number of digital ports, and fronthaul-efficient functional partitioning between the DU and RU. As a result, an architecture that appears attractive from a pure array-gain perspective may still face implementation bottlenecks unless the uplink processing chain is carefully redesigned for large-port operation.


\subsection{AI-RAN for X-MIMO}

The field-trial results also highlight the operational complexity of large-port X-MIMO systems, making them a natural candidate for AI-assisted RAN operation~\cite{c12}. In particular, 256-port transmission increases the MIMO dimension of CSI acquisition, beam management, scheduling, and fronthaul processing well beyond that of today's commercial systems. This creates a strong case for AI-RAN techniques such as AI-based CSI compression~\cite{c13}, channel prediction, beam selection, and cross-layer resource optimization~\cite{c14}. Another promising direction is site-specific learning of local propagation structure, which may help predict blockage, exploit recurring angular patterns, and reduce the control overhead associated with large-scale beamforming in practical deployments~\cite{c15}. More broadly, these research directions indicate that the evolution toward 7 GHz X-MIMO is not simply a matter of scaling antenna count, but of co-evolving architecture, signal processing, and operational intelligence so that large-port systems remain practically deployable.


\section{Conclusion}
\label{sec:conclusion}

This article examined the potential of 7 GHz X-MIMO from three complementary perspectives: architecture, system-level evaluation, and field validation. Our results show that scaling the BS architecture from today's 64-port class toward 256 digital ports can substantially strengthen both beamforming resolution and spatial multiplexing capability in the upper mid-band. Using a custom-built 256-port prototype, we further demonstrated the feasibility of 8-layer downlink SU-MIMO exceeding 3 Gbps over a 100 MHz channel in selected urban outdoor scenarios. Channel analysis based on measured data suggests that large-port arrays can exploit limited but usable spatial structure more effectively than conventional systems. At the same time, broader validation remains necessary for wider coverage conditions, uplink operation, mobility, and multi-user field performance. Overall, our results suggest that 7 GHz X-MIMO should be viewed not merely as an incremental extension of 5G, but as a compelling architectural evolution for unlocking the full potential of the upper mid-band for 6G.

\section*{Acknowledgment}

The authors would like to thank Keysight Technologies for their collaboration in this field trial. Keysight's high-capacity UE emulation platform played an important role in enabling the field tests in the 7 GHz band.

\end{document}